\documentclass[useAMS,usenatbib]{mn2e}
\usepackage{graphicx}
\newcommand{\ltt}{$\stackrel{<}{_\sim}$~}

\newcommand{\TAO}{$\tau_{\rm AqO}$~}
\newcommand{\TSL}{$\tau_{\rm d}$~}

\newcommand{\mum}{$\mu$m~}
\newcommand{\Martin}{Mart\'{i}n~}
\newcommand{\Magazzu}{Magazz\`u~}
\newcommand{\Bejar}{B\'{e}jar~}

\newcommand{\HHO}{H$_2$O~}

\newcommand{\Tef}{\mbox{$T_{\rm eff}~$}}
\newcommand{\Msun}{\mbox{\,M$_\odot$~}}
\newcommand{\vsini}{$v$~sin$i$~}

\title[Lithium in the brown dwarf LP\,944$-$20 revisited]{Lithium in LP\,944$-$20}

\author[
Ya.V.Pavlenko et al.]{Ya.V.Pavlenko$^1,^2$\thanks
{E-mail:yp@mao.kiev.ua (MAO)} \and H.R.A.Jones$^2$ \and
E.L.\Martin$^{3,4}$ \and E. Guenther$^5$  \and M.A. Kenworthy$^6$ 
\and M.R.Zapatero Osorio$^3$ 
 \\
\footnotemark[1]\thanks{Based on observations obtained on the European
Southern Observatory at Cerro Paranal, Chile in programs 68.C-0063(A) and
072.C-0110(B), the Anglo-Australian Telescope at Siding Springs Observatory
during commissioning observations for SPIRAL instrument; and the Keck  
Observatory in Mauna Kea, Hawaii.}
$^{1}$Centre for Astrophysics Research, University of Hertfordshire,
College Lane, Hatfield, Hertfordshire AL10 9AB, UK\\
$^{2}$Main Astronomical Observatory, Academy of Sciences of the Ukraine, Golosiiv
     Woods, Kyiv-127, 03680 Ukraine \\
$^3$ Instituto de Astrof\'\i sica de Canarias, La Laguna, Tenerife 38200, Spain \\
$^4$ University of Central Florida, Department of Physics, PO Box 162385, Orlando, FL 32816, USA
\\
$^5$ Th\"uringer Landessternwarte Tautenburg, 07778 Tautenburg, Germany
\\
$^6$ CAAO, Steward Observatory, 933 N.Cherry Ave, Tuscon, AZ 85721, USA }

\begin{document}
\date{}
\pagerange{\pageref{firstpage}--\pageref{lastpage}} \pubyear{2002}
\maketitle
\label{firstpage}
\begin{abstract}
We present a new estimate of the lithium abundance in the atmosphere
of the brown dwarf LP\,944$-$20. Our analysis is based on a
self-consistent analysis of low, intermediate and high resolution
optical and near-infrared spectra.  We obtain log N(Li) = 3.25 $\pm$
0.25 using fits of our synthetic spectra to the Li I resonance line
doublet profiles observed with VLT/UVES and AAT/SPIRAL. This lithium
abundance is over two orders of magnitude larger than previous
estimates in the literature. In order to obtain good fits of the
resonance lines of K I and Rb I and better fits to the TiO
molecular absorption around the Li I resonance line, we invoke a
semi-empirical model atmosphere with the dusty clouds located above
the photosphere. The lithium abundance, however, is not changed by the
effects of the dusty clouds. We discuss the implications of our
estimate of the lithium abundance in LP\,944$-$20 for the
understanding of the properties of this benchmark brown dwarf.
\end{abstract}
\begin{keywords}
stars: individual: LP\,944$-$20 --
           lithium abundance --
           stars: fundamental parameters --
           stars: late-type --
           stars: brown dwarfs --
           stars: evolution
\end{keywords}

\section{Introduction}

M dwarfs are of special interest to many branches of modern
astrophysics. The population of these numerous low-mass stars (M \ltt
0.6 \Msun) can contain an appreciable amount of the baryonic matter in
the Galaxy. The coolest M dwarfs (later than M6) could be young brown
dwarfs (\Martin, Rebolo \& Zapatero Osorio 1996) if they are young
enough.  The existence of brown dwarfs was predicted by Kumar (1963)
and Hayashi \& Nakano (1963).  The first unambiguous brown dwarfs
Teide 1 and Gliese 229B were reported by Rebolo, Zapatero Osorio \&
Mart\'\i n (1995) and Nakajima et al. (1995), respectively.

The ``Lithium test'' observation of the Li I resonance doublet at
$\lambda$ 670.8 nm was proposed by Rebolo et al. (1992) and developed
by \Magazzu et al. (1993) to distinguish young brown dwarfs from very
low-mass stars. This test has allowed the identification of many brown
dwarfs of spectral classes M and L (Rebolo et al. 1996;
\Martin et al. 1997, 1999; Kirkpatrick et al. 1999).

Due to the low temperatures in the cores of brown dwarfs of masses
$M<55M_{\rm Jup}$ lithium is not destroyed.
From $55M_{\rm
Jup}<M<75M_{\rm Jup}$, lithium is destroyed on timescales that can be
used to obtain nuclear ages for these very low-mass objects
(e.g. Chabrier et al. 2000).  However, the application of the lithium
test requires improvement of our knowledge about the physics of
atmospheres in low temperature regimes (Pavlenko et al. 1995).

LP\,944$-$20 (other names are APMPM J0340-3526, BRI B0337-3535, LEHPM
3451, 2MASS WJ0339352-352544) is an archetypical brown dwarf of
spectral type M9. A dim red dwarf-like star, Luyten Palomar (LP) 944-20 (M9 V)
was cataloged by Luyten \& Kowal (1975), and the SIMBAD database lists
more than  125 references for this object up to  May
2007. It is interesting to note that it is comparatively bright
($M_{\rm bol}$=14.22, Dahn at al. 2002) and also appears to be a young
object, a fact that is confirmed by the lithium detection of
Tinney (1998).

Using evolutionary tracks Tinney (1998) estimated the age of
LP\,944$-$20 to be between 475 and 650 Myr. Ribas (2003) found some
evidence for the membership of LP\,944$-$20 in the Castor moving group
in the solar vicinity ($R \sim 5-20~{\rm pc}$ from the Sun). The age of
LP\,944$-$20 as a member of the Castor group LP922-20 has been
estimated as $320 \pm 80~ {\rm Myr}$ (Ribas 2003).

An unexpectedly strong, solar like X-ray flare (log ($L_x/L_{\rm bol}$
= -4.1) of duration $\sim$ 2 hours was registered by {\it Chandra}
(Rutledge et al. 2000), but later XMM-Newton observations by \Martin \&
Bouy (2002) did not detect any X-ray activity. The quiescent X-ray
luminosity of LP\,944$-$20 is lower than that of the active Sun, and
X-ray flares on LP\,944$-$20 are very rare events in comparison with
more massive M-dwarfs (Hambaryan et al. 2004). However, LP\,944$-$20
is not the only M9 dwarf showing X-ray flares. Recently Stelzer (2005)
observed an X-ray flare (log ($L_x/L_{\rm bol}$ = -3.3) from the brown
dwarf binary with spectral types M8.5/M9 Gl569~B, which has an age of
about 300 Myr (Zapatero Osorio et al. 2004, see also Viti \& Jones
1999).

Our knowledge about the structure of the outermost layers of ultracool
dwarfs is far from complete (e.g. see discussion in Liebert et
al. 2003). Indeed, Berger et al. (2001) reported the discovery of both
quiescent and flaring radio emission from LP\,944$-$20. They observed
radio luminosities of 80 $\mu$Jy in quiescence and 2
$\mu$Jy at flare peak respectively, which are several orders of
magnitude larger than those predicted by the empirical relations between
X-ray and radio luminosities that has been found for late-type
stars. From the comparison of the continuous and flare fluxes at 8.5
GHz, Berger et al.(2001) concluded that the dwarf has a rather weak
magnetic field in comparison with flaring M-stars, and its emission
mechanism is the most likely non-thermal synchrotron radiation.

Tsuji et al. (1996) predicted ``dusty effects'' in the atmospheres of
cool dwarfs with \Tef $<$ 2600 K. The presence of large amounts of
dust in the highly dynamical atmospheres of fast rotating ultracool
dwarfs can yield ``weather phenomena'' (Jones \& Tsuji 1997, Gelino et
al. 2001; Mart\'\i n, Zapatero Osorio \& Lehto 2001; Clarke et
al. 2002; Bailer-Jones \& Lamm 2003, Caballero et al. 2003; Koen et
al. 2004).  LP\,944$-$20 also shows significant rotational broadening in
high-resolution spectra. The \vsini values determined by different
authors are in the range of 30 $\pm$ 2 km/s (Tinney \& Reid 1998;
Mohanty \& Basri 2003, Guenther \& Wuchterl 2003; Jones et al 2005;
Zapatero Osorio et al. 2006). The rotation period has not been
determined yet, but photometric variability was reported by Tinney \&
Tolley (1999). These surface inhomogeneities could be due to magnetic spots
or dusty clouds modulated by stellar rotation.


Tinney (1998) reported the observation of a strong
lithium line in the spectrum of LP 944$-$20 as evidence of
its youth and substellar nature, and estimated a lithium
abundance of log N(Li) = 0.0 $\pm$ 0.5, indicating it had
depleted an important amount of  lithium.
This is so far the only
determination of the lithium abundance in LP\,944$-$20, which was
obtained via comparison of the observed feature width with theoretical
curves of growth.  However, the observed atomic line at 670.8 nm
is measured with respect to the molecular background formed by TiO
lines and therefore a detailed spectral analysis needs to be made.
Here we present new computations using different approaches and model
atmospheres indicating that LP\,944$-$20 has not depleted any lithium.

This paper is organized as follows: Section 2 presents the
spectroscopic data of LP\,944$-$20 available in the
literature. Section 3 describes the procedure followed to compute
synthetic spectra and spectral energy distributions  using
standard model atmospheres. We fit these theoretical spectra to the
observations.  In Section 4 we discuss the results obtained in the
framework of the conventional approach and we derive a new estimate of
the lithium abundance in LP\,944$-$20. In section 5 we investigate the
impact of the possible deviations of physical parameters on our
results. We propose a semi-empirical (SE) model for the formation of
molecular features and strong resonance lines of alkali elements in
the spectrum of LP\,944$-$20.  We present the results of applying the
SE model for fitting K I, Rb I and Li I lines in high-resolution
spectra and spectral energy distributions. We derive a new estimate of
the lithium abundance in LP\,944$-$20 with our SE model atmosphere and in
Section 6 we discuss the implications of our results.

Coloured versions of plots are available at
ftp://star-ftp.herts.ac.uk/pub/Pavlenko/lp944.20 

\section{Observations}
 The observed spectra used for this paper have been published elsewhere. 
In order to simplify the text of the paper we adopt a labelling 
syntax for them as follows:
\begin{itemize}
\item CASPEC: optical spectra obtained by Tinney \& Reid (1998) 
  using the Cassegrain Echelle Spectrograph on the ESO 3.6 m
  telescope (R $\sim$ 18000). The infrared part of the spectral
  energy distribution was obtained by UKIRT (see references in Leggett
  et al. 2001). The combined spectrum was taken from the Leggett's
  database \\ (ftp://ftp.jach.hawaii.edu/pub/ukirt/skl/).\\  We believe
  Leggett et al. (2001) normalised the spectral data using 
  broad-band photometry but the error on this is of the order of
  10\%.  We have rescaled the flux
  beyond 1 $\mu$m by 10\%~to improve the match between observed and
  synthetic spectra.
\item SPIRAL: spectra of medium resolution obtained with SPIRAL Phase A
fibre-fed spectrograph on
  the AAT. The FWHM was typically 2.4 to 2.7 pixels over the whole data,
  the GFWHM, i.e.  FWHM of the smoothing Gaussian, was of order 0.883
  \AA~ for the Li I 670.8 nm line (R=7500). See Kenworthy et
  al. (2001) for the data reduction details.
\item NIRSPEC:  the near-infrared spectrum of LP\,944$-$20
  (1.240--1.258 \mum, R $\sim$ 22000) was taken with Keck II.  All
  details about those observations have been published in Zapatero
  Osorio et al. (2006), and \Martin et al. (2006).  
\item UVES: high spectral resolution spectrum obtained with UVES
  (Ultraviolet-Visual Echelle Spectrograph) on the VLT Unit telescope
  2 (KUEYEN) at Paranal in service mode. The setting simultaneously
  covers the wavelength regions from 667.0 to 854.5 nm.  We used the
  average of 15 individual spectra obtained for the purpose of radial
  velocity monitoring.  Details of the data reduction and calibration
  can be found in Guenther \& Wuchterl (2003).  The theoretical
  resolution of our UVES spectra is of order 40000, corresponding to
  the GFWHM = 0.17 \AA.
\end{itemize}

\section{Standard spectral modelling}

To determine the surface temperature ($T_{\rm eff}$) of LP\,944$-$20
we have used the equation $L = 4 \pi R^2 \sigma T_{\rm eff}^4$,
normalized to solar units. The trigonometric parallax of LP\,944$-$20
is found to be 200.7$"$ $\pm$ 4.2$"$, which is the mean value of the
measurements obtained by Tinney (1998) and Dahn et al. (2002), yielding
a distance of 5.0\,pc to the dwarf. By using the bolometric
corrections for the $J$- and $K$-bands and the observed near-infrared
photometry of LP\,944$-$20 available in the literature, we have
derived its luminosity at log $L/L_\odot$ = $-$3.79\,$\pm$\,0.03,
which is in agreement with the previous determination by Dahn et al.
(2002). The error bar accounts for the uncertainties in the
photometry, bolometric corrections and parallax.

The radii of ``old'' ($\ge$500 Myr) brown dwarfs are nearly
independent of mass and age, with a mean radius of 0.09~$R_{\odot}$
(see Burgasser 2001) we derive $T_{\rm eff}$ =
2170~K.  However, the fact that lithium is indeed detected in the
optical spectrum of LP\,944$-$20 indicates that this object is
significantly younger than other field dwarfs of similar
classification, which do not show the lithium feature in
absorption. Tinney (1998) and Ribas (2003) state that the age
of LP\,944$-$20 is in the range 240--650~Myr, with a likely value at
320~Myr. For such a young age, recent evolutionary models predict
brown dwarf (0.05--0.07~$M_\odot$) radii of about 0.10~$R_{\odot}$
(Chabrier \& Baraffe 2000; Burrows et al. 1997), i.e.~11\%~larger than
typical field brown dwarfs. This yields a temperature $T_{\rm eff}$ =
2040~K, which is the value that we will use throughout the present
paper. Recently, Bihain et al. (2006) have shown that proper motion
Pleiades brown dwarfs may have radii similar to those of much older
substellar objects, suggesting that brown dwarfs collapse faster than
expected. To account for this, we estimate the uncertainty in our
$T_{\rm eff}$ determination of LP\,944$-$20 to be $\pm$150~K, which
includes the luminosity error bar and the various object sizes valid
for an age interval 120--1000~Myr.

Our strategy is divided into the following steps: {\sl (i)}
Theoretical spectra are computed and fitted to the observed spectral
energy distribution of LP\,944$-$20 from 0.65 to 2.5 \mum. {\sl
(ii)} We then compare our synthetic spectra  obtained following a
standard classical approach with some  atomic and molecular
features in the observed spectrum (in particular, the resonance lines
of K I, Rb I). Here we use spectra of medium and high spectral
resolution.  Finally, {\sl (iii)} we determine the
sensitivity of our best fit lithium abundance to the input parameters
of our models.

\subsection{Synthetic spectra}

Theoretical spectral energy distributions (hereafter we use the term
``synthetic spectra'') were computed for model atmospheres of dwarfs
with effective temperatures \Tef = 1800--2400 K and log $g$~ = 4.0 --
4.5  (cm s$^{-2}$) from the DUSTY and COND model atmospheres
(Allard et al. 2001) of solar metallicity (Anders \& Grevesse 1989).
Hereafter we use the model atmosphere notation 2400/4.5/0.0 to mean
\Tef = 2400 K, log $g$~ = 4.5, [M/H] = 0. We  will focus on
results obtained with 2000/4.5/0 model atmosphere  because as
previously indicated, 2040 K is an appropriate effective temperature of
LP\,944$-$20, and log $g$~ = 4.5 since LP\,944$-$20
is considered to be younger than field dwarfs of similar
classification. Nevertheless, computed spectra for other temperatures and
gravities were used to study the dependence of our results on these
parameters.

The dominant opacity sources in the optical and infrared spectra of
LP\,944$-$20 are absorption by band systems of diatomic molecules,
such as TiO and VO. Computations of synthetic spectra were carried out
by the program WITA5 (Pavlenko et al. 2000) assuming LTE, hydrostatic
equilibrium for a one-dimensional model atmosphere and without sources
and sinks of energy. The equations of ionisation-dissociation
equilibrium were solved for media consisting of atoms, ions and
molecules. We took into account $\sim$ 100 components (Pavlenko
1998a). The constants for the equations of chemical balance were taken
mainly from Tsuji (1973) and Gurvitz et al. (1989).

\begin{figure}  
\includegraphics[width=88mm]{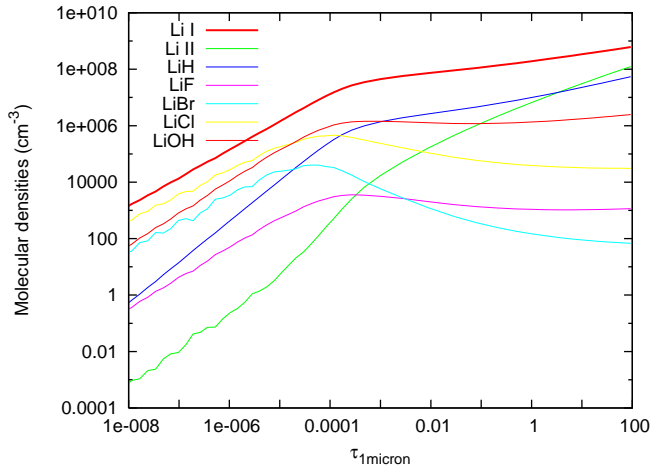}
\caption[]{\label{_lidens}
Molecular densities of lithium containing species in the 2000/4.5/0.0
model atmosphere (\Tef = 2000 K, log $g$ = 4.5, solar metallicity).}
\end{figure}

Our lithium containing species list includes Li I, Li II, LiOH, LiH,
LiF, LiBr, LiCl. Our computations show that within our (\Tef, log $g$)
range neutral lithium dominates across the atmospheres - only 5\%~of
lithium atoms are bound in molecules in the outermost layers of
atmosphere (Fig. \ref{_lidens}). The most abundant lithium containing
species are molecules of LiCl (in the outermost layers) as well as Li
II and LiOH (in the photospheric layers).

\subsection{\label{_molads} Molecular and atomic lines opacity}

The molecular line data are taken from different sources: TiO line
lists by Plez (1998); CN lines from CDROM 18 (Kurucz 1993); CrH and
FeH lines from Burrows et al. (2002a) and Dulick et al.(2003),
respectively. Atomic line list data are taken from VALD (Kupka et
al. 1999).

\begin{figure}   
\includegraphics[width=88mm]{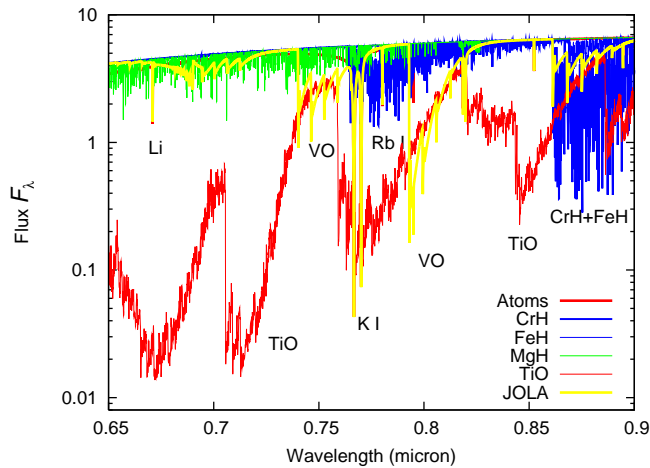}
\caption[]{\label{_ident}
A plot identifying the main
features in a 2400/4.5/0.0 model spectrum, showing
their relative strengths.}
\end{figure}

The profiles of molecular and atomic lines are determined using the
Voigt function $H(a,v)$ except for the strong resonance doublet lines of
Na I (0.5891, 0.5897 \mum~) and K I (0.7667, 0.7701 \mum~).
(see the section \ref{_kna_}). The parameters of natural
broadening $C_2$ and van der Waals broadening $C_6$ of absorption
lines are taken from Kupka et al. (1999), or in their absence computed
following Uns\"old (1955). In our computations we used a correction
factor $E$ for van der Wals broadening of atomic lines.  Parameter $E$
describes differences in broadening parameters computed in the
framework of classical and quantum physics approaches. For resonance
lines of alkali metals $E$ = 1--2 (see Andretta et al. 1991).
Owing to the low electron and ion densities in low temperature M dwarf 
atmospheres, Stark broadening may be neglected.
In
general the effects of pressure broadening prevail. Computations for
synthetic spectra to be fitted to observed spectra across the 0.65 - 0.9 \mum
are carried out at intervals of 0.5 \AA. For comparison with high
resolution spectra we computed spectra with an interval of 0.02 \AA. The spectrum
broadening is modelled by Gaussian plus rotational profiles set to the
resolution of the observed spectra. Rotational broadening was taken
into account following the Gray (1976) formulae with \vsini = 30
km/s (see Jones et al. 2005). VO band opacity is computed in the frame
of the JOLA approximation (see Pavlenko et al. 2000 for more
details). The relative importance of the different opacities
contributing to our synthetic spectra is shown in Fig. \ref{_ident}.

In this paper we  also pay careful attention to the modelling of
the rubidium resonance line  at 780.2 nm. Resonance lines of Rb I
have superfine structure (Lambert \& Mallia 1968, Lambert \& Luck
1976), which is important to account when considering weak absorption
lines, but the Rb I lines in our spectra are strong. Moreover, they
are broadened by rotation and turbulent motions and so the superfine
splitting of Rb I lines can be neglected (see ibid.).  We assume that
all atoms of rubidium are the $^{58}$Rb isotope.  Most features
of the observed spectra should be fitted using the same atmospheric
parameters of temperature and gravity. We will use our fits to the Rb
I line as a test of the consistency of our spectroscopically derived
results to log $g$ and lithium abundance in the atmosphere of
LP\,944$-$20.

\begin{figure*}  
\includegraphics[width=88mm]{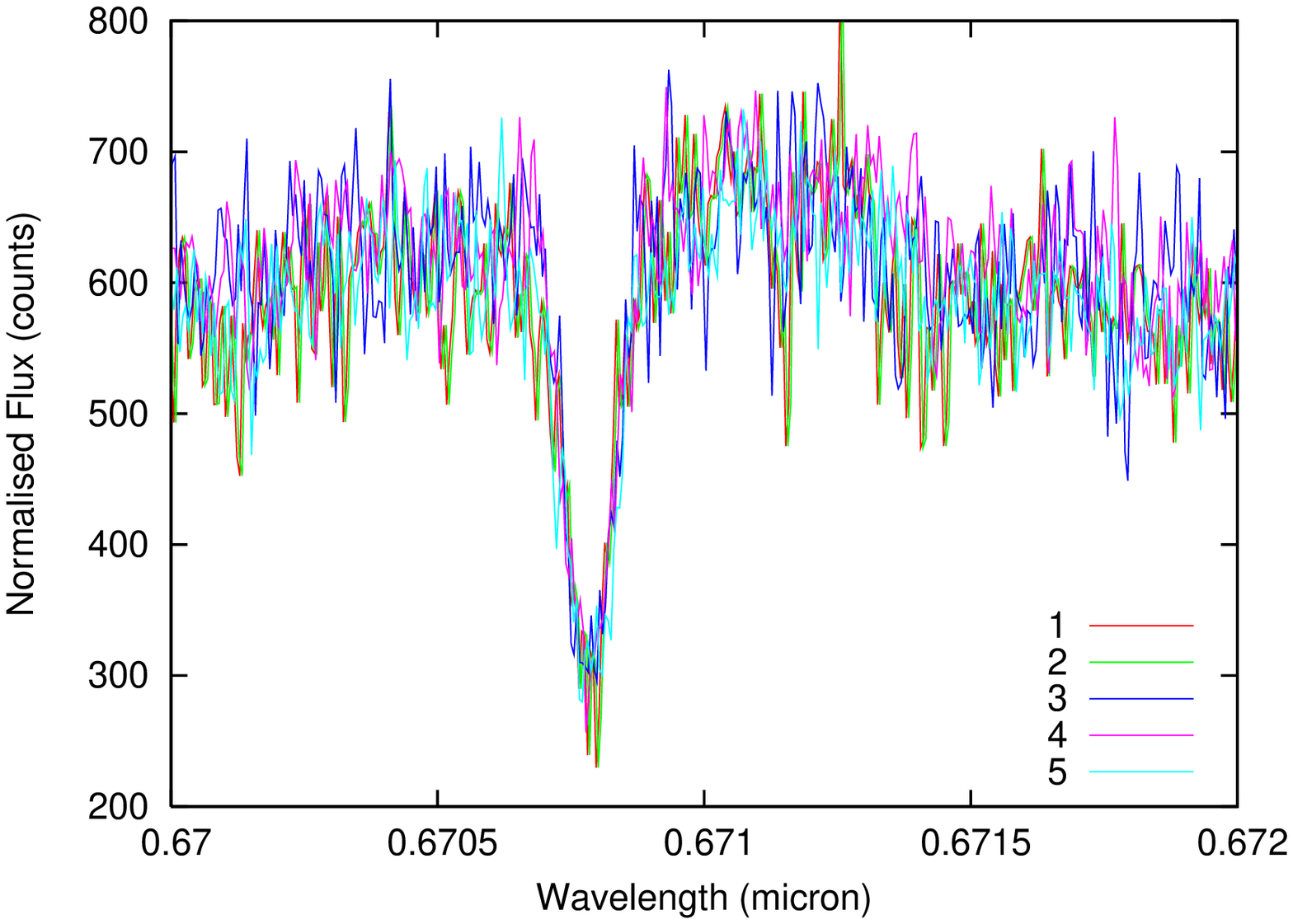}
\includegraphics[width=88mm]{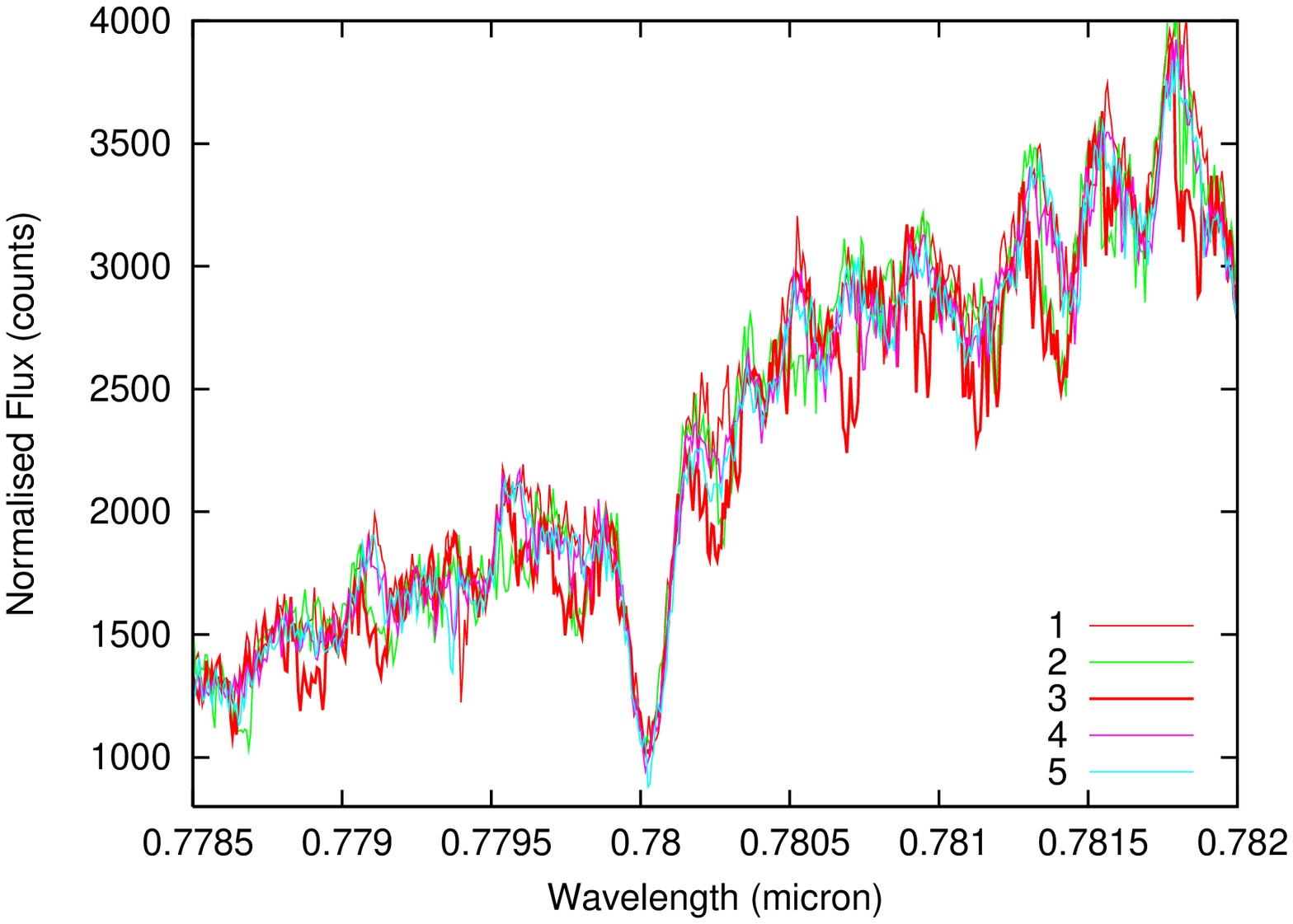}
\caption[]{\label{_LiRb} Observed fluxes of LP\,944$-$20 for
the first five (in chronological order) UVES echelle orders across
spectral regions containing Li I and Rb I resonance lines.}
\end{figure*}

\subsection{\label{__liobs}Observed profiles of Li I and Rb I}

The flux variations across Li I and Rb I line regions obtained in a
few separate observing runs are shown in Fig.  \ref{_LiRb}. We
show there a sequence of the first 5 observed fluxes from a set of 15
spectra which were used to get the combined spectrum used in our
analysis. It is worth noting that changes in the Rb I line 780.2 nm
profile are rather small.  However, we see here some variations in
molecular background formed by the haze of TiO lines. In general, the 
response of the TiO molecular densities to changing temperatures in
the line forming region should be much higher in comparison with
changes of the ionisation equilibrium of alkali metals (see Pavlenko
1998).  Some variations of the observed fluxes across the
780 nm region are real. Analysis of the topic is beyond the scope
of our paper.

For the 670 nm region, we find that the data from individual 
UVES epochs appear to be reasonably fit, albeit at low
signal-to-noise, by the same parameters as the higher signal-to-noise
Rb lines. However, when the dataset is combined (15 low
signal-to-noise epochs spread over more than a year) a considerably
broader  Li I profile results.  This is not due to  radial
velocity changes for which this dataset has already being 
corrected to a level of the order of 80 m\,s$^{-1}$ (Guenther \&
Wuchterl 2003). This apparently changing line profile is presumably
due to  either variability and/or relatively low signal-to-noise
ratio of the UVES data at these wavelengths and leads us to fit the
combined profile with a lower effective resolution. We find the
lithium region to be best fit with \vsini = 32 km/s and a GFWHM = 1
\AA.

\subsection{\label{_kna_}K and Na resonance lines}

In the spectra of ultracool dwarfs the resonance lines of Na I and K I
are very strong. They govern the spectral energy distributions of
L-dwarfs across a wide spectral region (see Pavlenko 2001 for more
details). Their formally computed equivalent widths may be of the
order of a few thousand \AA. For these features we cannot use a
classical collisional approach to compute the wings of these
superstrong lines. In the dense, cool atmospheres of late M, L and T
dwarfs the pressure broadening of K I and Na I can be computed with a
quantum chemical approach (Burrows \& Volobuyev 2003; Allard at
al. 2003).

For this work we use potentials of quasi-stationary chemical
interactions of atoms K and Na with the most numerous species, i.e.,
atoms He and molecules H$_2$ computed by GAMESS (Granovsky et
al. 1999). Our procedure is described in more detail in Pavlenko et
al. (2007). Here we point out that in computations of K I profiles we
used a combined profile: cores of these lines were computed in the
frame of the collisional approach, and their wings ($\delta\lambda >$
40 \AA) were treated by quasi-stationary theory. It is worth noting
that K and Na lines have comparatively weak wings in our range of \Tef
and log $g$.  Thus, in observed spectra we see only cores of strong
absorption lines against a background of molecular bands.

For other alkali lines we used a conventional, i.e., collisional theory
of pressure broadening, using van der Waals formulae for computation
of damping constants described in Pavlenko (2001). The relative
strength of the resonance lines of neutral alkali lines is shown in
Fig. \ref{_ident}.

\subsection{\label{eq}Fit to the observed spectral energy distribution}

To get the best fits of our theoretical spectra to the observed
spectral energy distribution we followed the scheme described in
Pavlenko et al. (2006)  and references therein.  Namely, we find
the minimum value of the following equation:
\begin{equation}
S(f_{\rm h},f_{\rm s},f_{\rm g})=\Sigma(f_{\rm h} \times H^{\rm synt}-H^{\rm obs})^2
\end{equation}
Here $H^{\rm obs}$ and $H^{\rm synt}$ are observed and computed
fluxes, $f_{\rm h}$ is a normalisation parameter,  $f_s$ is the
relative wavelength shift of the observed spectra, and $f_g$ is
related to the broadening parameters. In our case, we applied both
rotational and instrumental broadening. These were modelled by Gray
(1996) formulas (rotational broadening) and Gaussian functions (instrumental
broadening).

\begin{figure}   
\includegraphics [width=88mm]{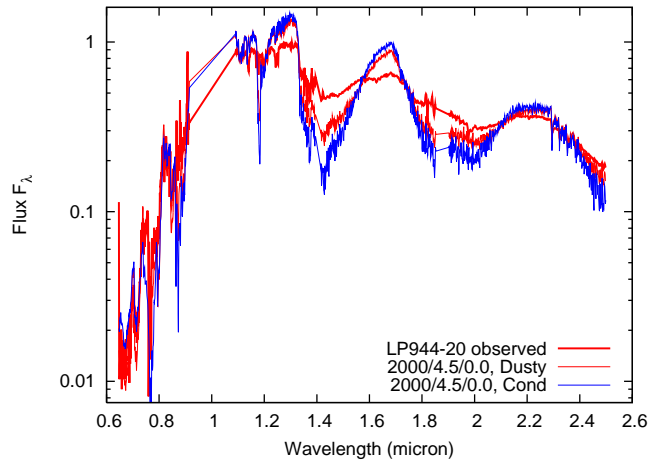}
\caption[]{\label{__fitA0} Fits of the synthetic spectra computed for
DUSTY and COND model atmospheres 2000/4.5/0.0 to the observed spectral
energy distribution of LP\,944$-$20. There is no data between 0.9
and 1.1 $\mu$m.}
\end{figure}

\begin{figure}  
\includegraphics [width=88mm]{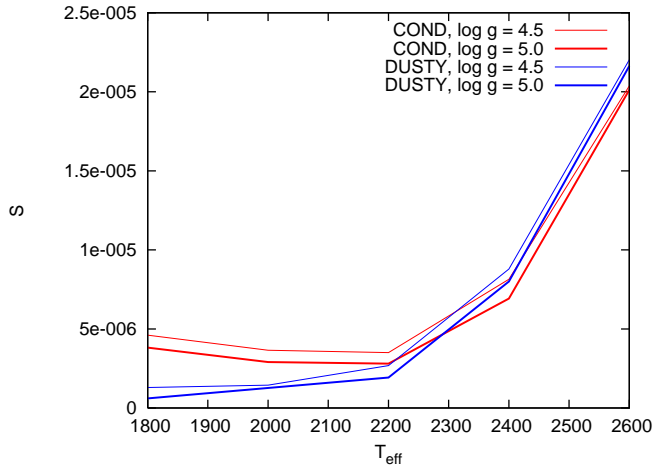}
\caption[]{\label{__As}
Values of min ($S$) computed for a grid of DUSTY and COND
model atmospheres in the ``standard'' framework.}
\end{figure}

\section{Results based on standard spectral modelling}
\subsection{\label{__fitto}Fit to the observed spectral energy distributions}

In Fig. \ref{__fitA0} we show fits to the observed spectral energy
distributions (SEDs) of LP994-20 across the 0.65--2.5 $\mu$m region.
We carried out our fits for the DUSTY and COND model atmospheres. We
exclude from the fitting the spectral region 0.86--1.96 $\mu$m 
because there is a gap in the observed spectrum, and strong spectral
features formed by absorption of \HHO and other molecules at these
wavelengths cannot be fitted properly even for the objects of higher
\Tef (Pavlenko et al. 2006). Nonetheless, our approach allows us to
use the optical region governed by TiO bands where flux is very
sensitive to \Tef, and the region beyond 1.96 $\mu$m dominated by \HHO
and CO bands. Line lists of these molecules for the latter region is
of high quality (Jones et al. 2003; Pavlenko \& Jones 2003). In
general, fits of our theoretical spectra computed with DUSTY model
atmospheres to the observed spectra look better. In Fig. \ref{__As} we
show values  of min $S$ computed for a grid of DUSTY and COND model
atmospheres. It is worth noting:
\begin{itemize}
\item  The values of $S$ are smaller for the DUSTY models
across a wide range of effective temperature and gravity,
particularly below 2300 K.
\item We get min $S$ at \Tef = 2000--2200 K for the COND models. 
This result derived through spectral analysis is consistent
with our estimation of the surface temperature for LP\,944$-$20
obtained in Section 3 using radii predicted by substellar evolutionary
models, the object's astrometry and broad-band photometry.
\item  Temperatures in the outermost layers of the DUSTY models  are
higher in comparison with the COND models. As s result for the DUSTY models, 
$S$ appears to decrease even for \Tef $<$ 2000K. 
\end{itemize}

 We remark that any spectral analysis is affected by uncertainties
associated with the adopted structure of model atmospheres, molecular
opacities, and non-classical effects like the presence of a
chromosphere, veiling, etc. Additionally, the fits shown in
Fig. \ref{__fitA0} do not provide a good match to the observed SED
of LP\,944$-$20. Therefore, these results should be understood as a
qualitative confirmation of the \Tef that we have
assumed for LP\,944$-$20. From Fig. \ref{__fitA0}, it becomes apparent
that additional parameters describing the atmosphere of LP\,944$-$20
are needed to obtain a better reproduction of the observed data. We
will further discuss on this in Section 5. Next, we will show the
results for the high resolution ``standard'' analysis.

\subsection{Resonance lines of K I and Rb I}

\begin{figure}  
\includegraphics [width=88mm]{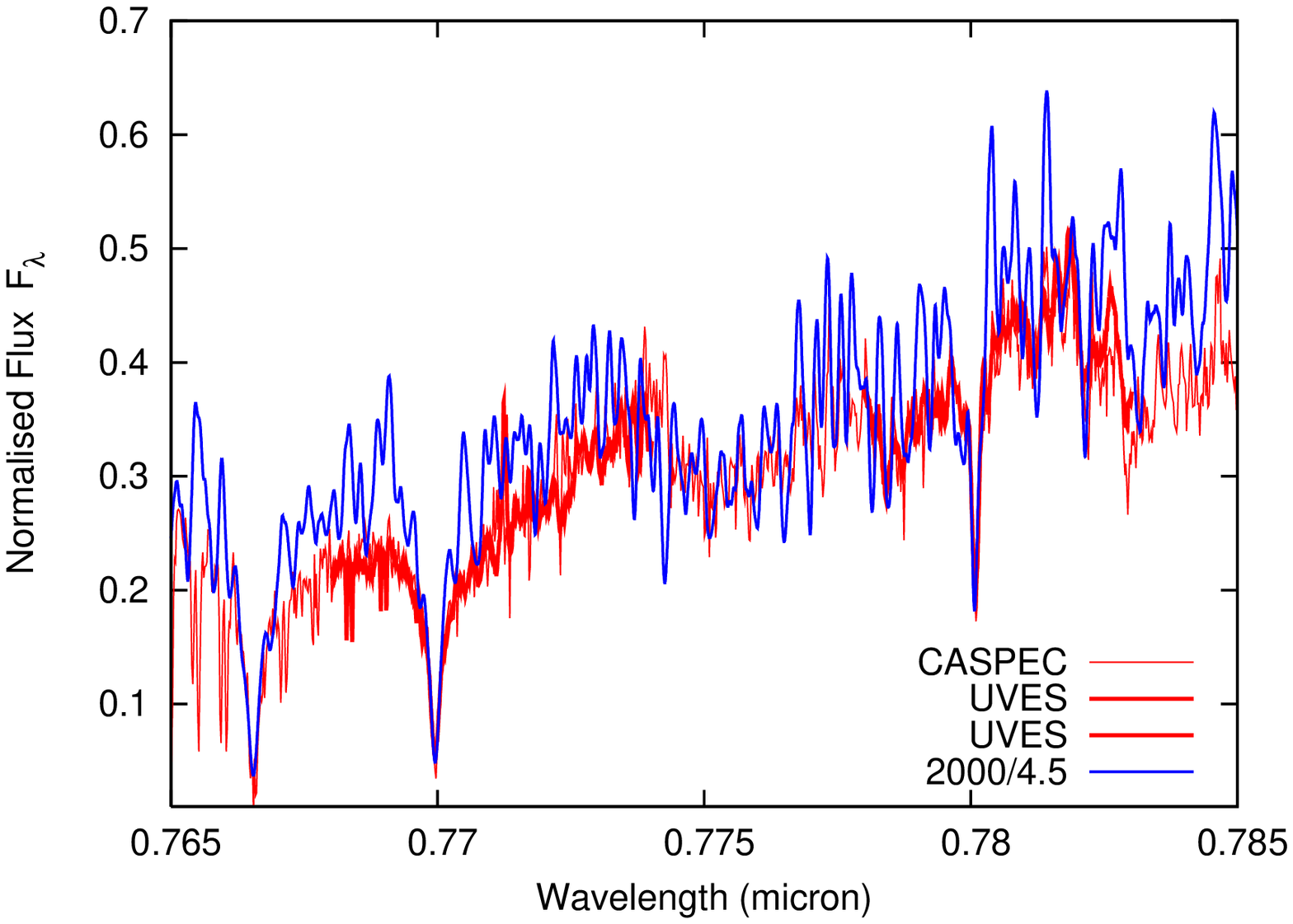}
\includegraphics [width=88mm]{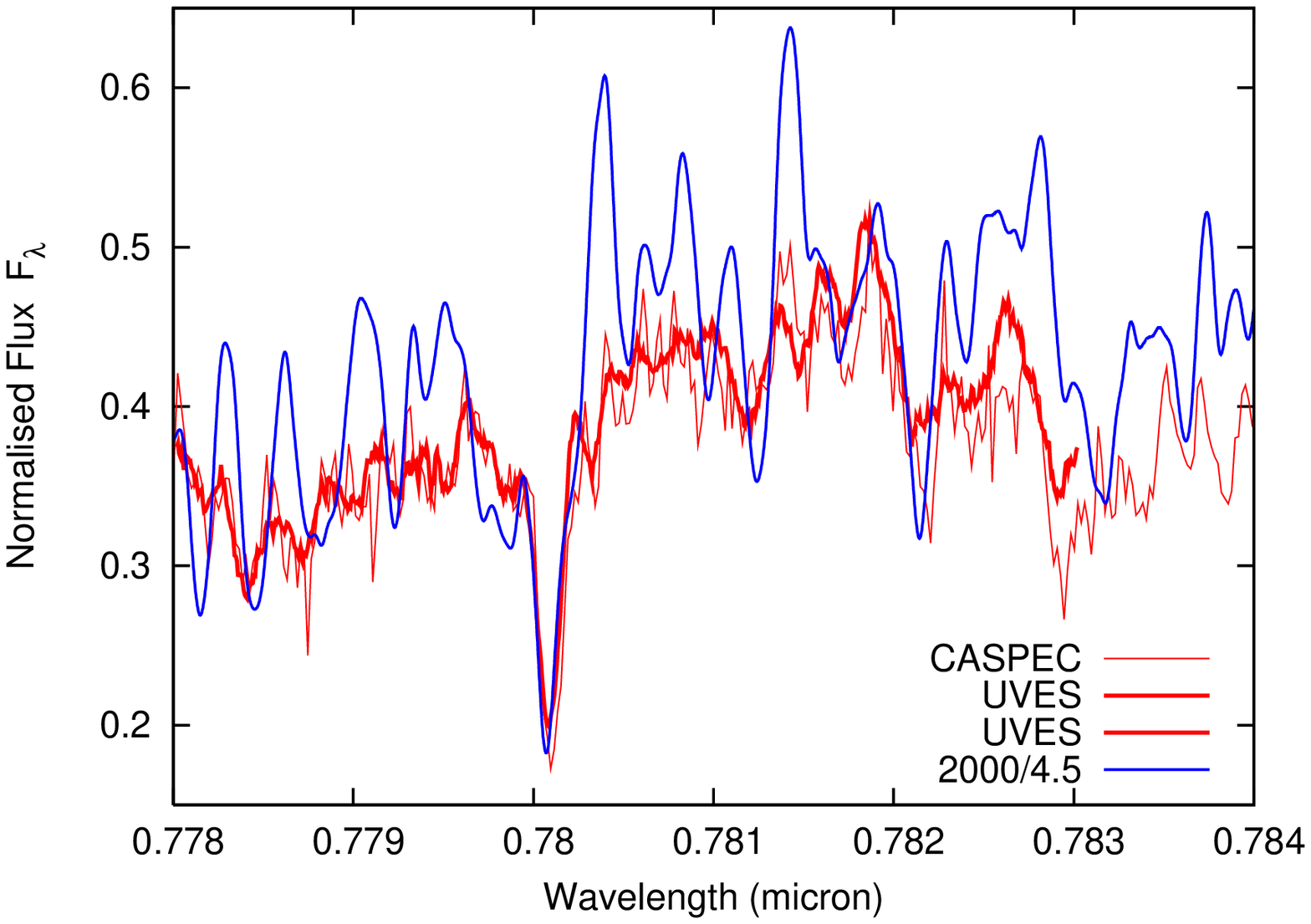}
\caption[]{\label{__KRbaa} Fits to the K I and Rb I resonance lines
observed in the spectra of LP\,944$-$20 following the ``standard''
approach. The bottom panel shows an enlargement around the Rb I
line. Theoretical spectrum has solar metallicity.}
\end{figure}

Resonance doublets of K I and Rb I are recorded 
simultaneously. This provides the opportunity to study these lines at
a single epoch.   The comparison of CASPEC, SPIRAL and UVES
spectra shows all these data agree well in spectral slope, widths and
intensities of lines. The fits to K I and Rb I resonance line are
shown in Fig. \ref{__KRbaa}. The Rb I line is more suitable for
analysis, because it is less saturated  than the KI doublet. We
can fit our synthetic spectra to the observed Rb I profile taking into
account the rotation velocity \vsini = 30 km/s and GFHWM = 0.17 \AA~
(see the bottom panel of Fig. \ref{__KRbaa}). Note that other
features mainly due to molecular absorptions of TiO and VO are not so
well reproduced in intensity by the models. Nevertheless, the
\Tef = 2000 K and log $g$ = 4.5 model nicely fits the K I and Rb I
resonance profiles. We note that higher gravities do not provide a
better match to the observed profiles of the akalis (they predict
stronger lines), as we will show next.

Rubidium in LP\,944$-$20 was also studied by Basri et
al. (2000). These authors derived \Tef = 2600 K from the spectral
fitting of the Rb I resonance lines. 
However, Basri et
al. (2000) point out that their spectral analysis of the Rb I lines
indicates higher temperatures than their analysis of
caesium lines (2400 K). 
Our computations for 2400 K do not
provide fits to the SED of LP\,944$-$20 as acceptable as those
computed for 2000--2200 K (see Fig. \ref{__As}). LP\,944$-$20 is
classified as an M9V field brown dwarf in the literature, thus it
shares the same spectral classification as LHS\,2924, which was also
included in the sample of Basri et al. (2000). These authors found
that LP\,944$-$20 is cooler than LHS\,2924 in agreement with the more
recent results of Dahn et al. (2002), who determine \Tef = 2367 and
2138 K for LHS\,2924 and LP\,944$-$20, respectively. This value is in
much agreement with our adopted \Tef for the young brown
dwarf. We note that our synthetic spectra provide
reasonable fits to the observed profiles of all alkali lines in
LP\,944$-$20 using the same atmospheric parameters, i.e., without the
need of different values of temperature and gravity for the various
lines present in the spectra.

\subsection{Near-infrared doublet of K I}

\begin{figure}  
\includegraphics [width=88mm]{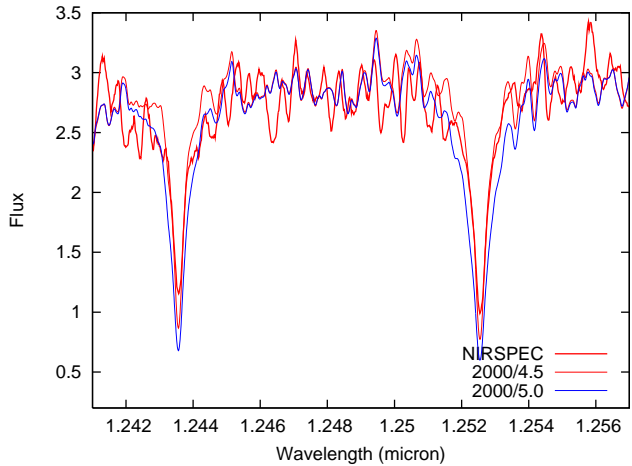}
\caption[]{\label{__IRa} Fits of DUSTY synthetic spectra (solar
metallicity, ``standard'' approach) to the observed infrared K I
subordinate doublet lines.}
\end{figure}

 We have also fit the NIRSPEC spectrum of LP\,944$-$20, which
contains the subordinate lines of K I at 1.2432 and 1.2522
$\mu$m. These lines turn out to be gravity-sensitive features as
observationally shown by McGovern et al. (2004). Their excitation
potentials are rather low ($\sim$ 1.610 eV), thus they can be used for
the analysis of stellar spectra. Our fits to the observed profiles of
K I subordinate doublet using the standard modelling procedure
are shown in Fig. \ref{__IRa}.  The molecular background at these
wavelengths is dominated by absorption of \HHO; CrH and FeH contribute
to the opacity as well.  Unfortunately, good quality
line lists for CrH and FeH do not exist.

Nonetheless, we get a good fit to the observed profiles of these lines
(Fig \ref{__IRa}) using \Tef = 2000 K and log $g$ = 4.5. Note
that higher gravity models deviate significantly from the
observations. From the spectral analysis of young M-type objects
carried out by Mohanty et al. (2004), these authors pointed out that
their derived gravities for the coolest M dwarfs deviate significantly
up to 0.75 dex from the isochrone predictions in the sense that
measurements are smaller. Our gravity derivation for LP\,944$-$20 also
appears to be lower than the prediction of log $g$ $\sim$ 5.0 based on
the evolutionary models by Baraffe et al. (1998) and the likely age of
the brown dwarf.

 While the overall profiles of the near-infrared K I
lines are reasonably reproduced by our computations, the cores of the
lines in the theoretical spectra are deeper than those in the observed
spectra. These differences in the intensity of the line cores of the
saturated lines  may be explained by  non-local
thermodynamical equilibrium (NLTE) effects or chromospheric
effects.  However, other possible physical explanations can be
found, especially those related to the depletion of refractory
elements into condensates in cool atmospheres (see Section 5).

\subsection{\label{__ll} Lithium lines and abundances}

\begin{figure}   
\includegraphics [width=88mm]{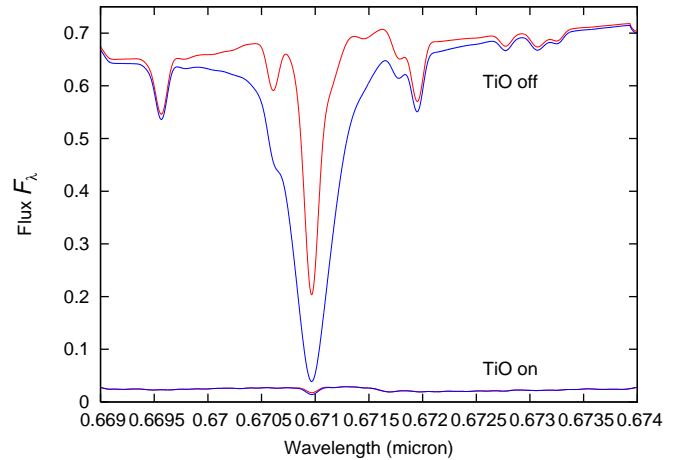}
\caption[]{\label{__Liaa} Theoretical profiles of atomic lines in the
  670.8 nm region computed both with and without TiO
  absorption in the 2000/4.5/0.0 model atmosphere. The
  strongest line is due to Li I.  The vacuum scale of wavelengths is
  used.  Synthetic spectra are convolved with GFWHM = 1~\AA~ and
  \vsini = 30 km/s.  Solid and dashed lines show spectra computed for
  log N(Li) = 2.0 and 3.2, respectively.}
\end{figure}

The relative strength of atomic lines around the Li resonance doublet
and TiO bands are shown in Fig. \ref{__Liaa}. These computations were
carried out for two cases  to show the dependence of the strength of
the Li I line on the presence of TiO: {\sl (i)} in the first case we
only take into account the absorption for atomic lines, and {\sl (ii)} in the
second and more realistic case we also accounted for molecular
absorption. While in the former case, analysis of atomic lines can
be carried by equivalent width studies, only pseudoequivalent widths
(pEWs), i.e., equivalent widths (EWs) measured in respect to the local
psudocontinuum formed by molecular and atomic line background, can be
obtained from the spectra computed in case {\sl (ii)} (see also
Pavlenko 1997).

\begin{figure}   
\includegraphics [width=88mm]{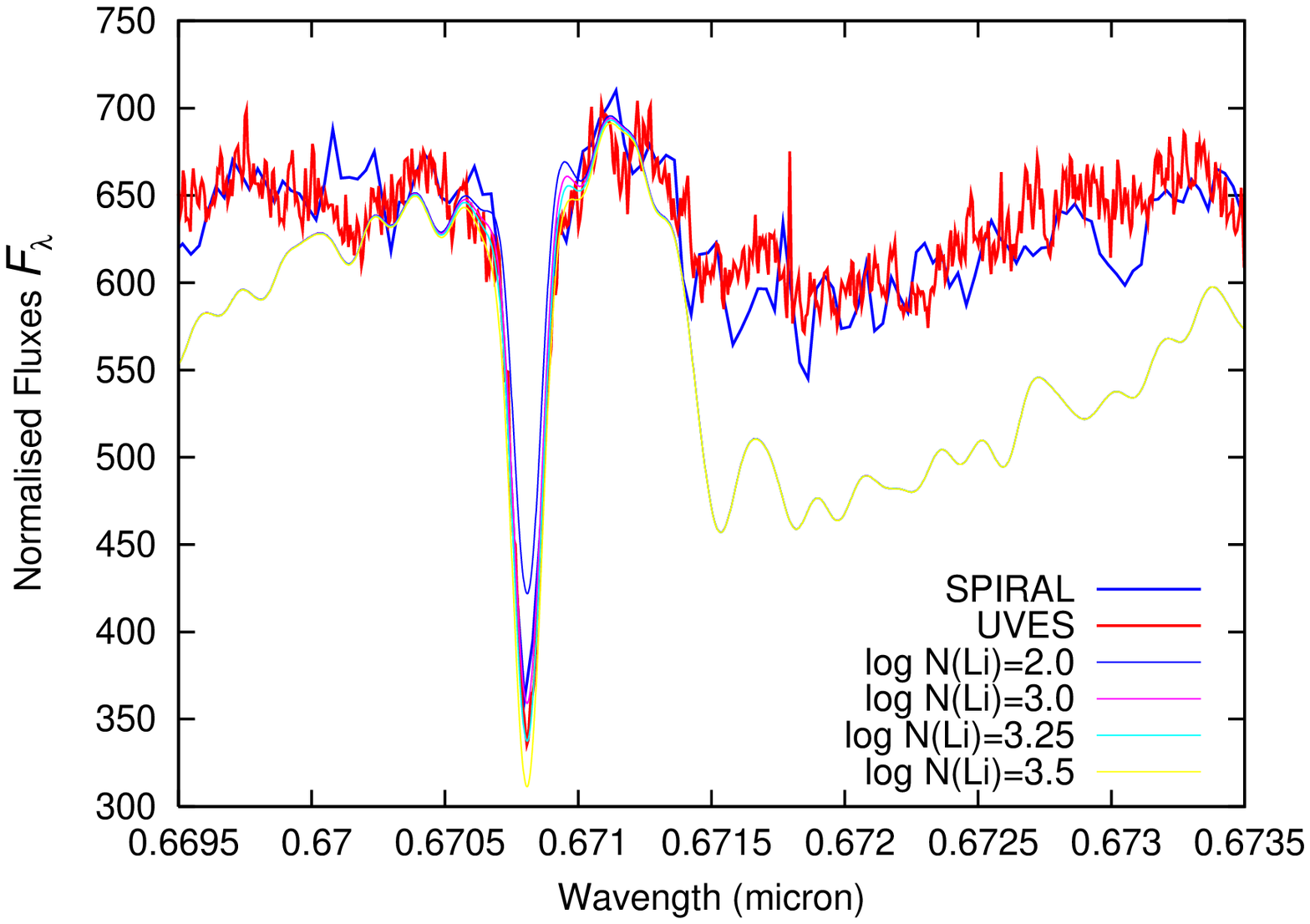}
\includegraphics [width=88mm]{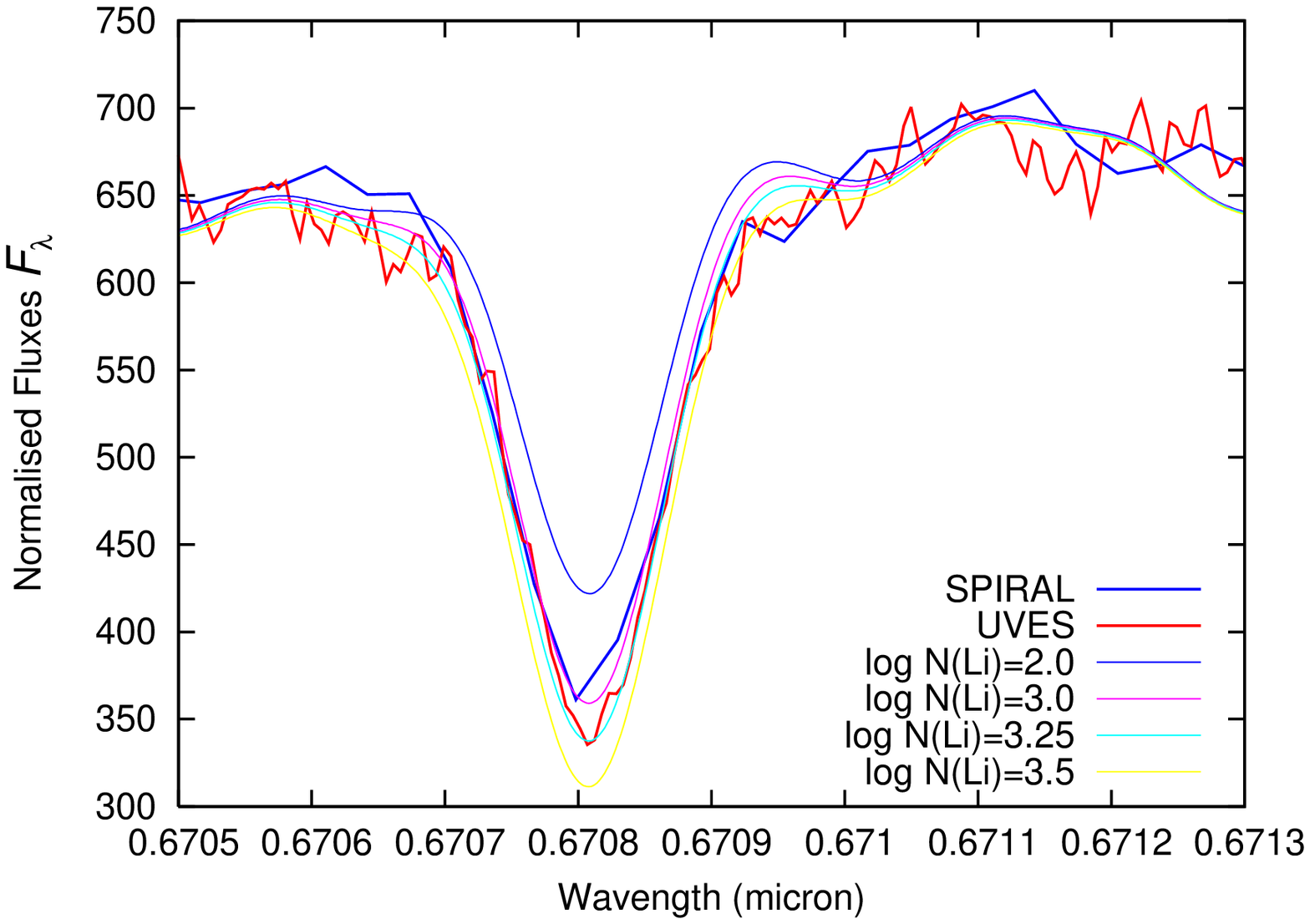}
\caption[]{\label{__Liba} Fits of the lithium 670.8 nm resonance
doublet computed for DUSTY model atmosphere 2000/4.5/0.0 and different
lithium abundances in the ``standard'' framework to observed spectra
of LP\,944$-$20. The bottom panel shows an enlargement around the
lithium atomic line.}
\end{figure}

Equivalent and pseudoequivalent widths of the Li I resonance doublet
differ significantly (see Zapatero Osorio et al. 2000; Pavlenko 2005).
The computed EW of the Li I feature at 670.8 nm for the DUSTY model
atmosphere 2000/4.5/0 and log N(Li) = 0.0 is 0.54 and 0.58 \AA~ for
turbulent velocities of 2 and 3 km s$^{-1}$, respectively. For log
N(Li) = 3.25 (i.e., cosmic abundance) we obtain  much larger
EW of 13.5 \AA~(this theoretical measurement accounts for the
broad wings of the doublet $\pm$ 40 \AA~ from its
core). Interestingly, the formally measured pEW of the lithium line
obtained for log N(Li) = 3.25  and the synthetic spectra of case
{\sl (ii)} is considerably smaller, much in agreement with the
observations of LP\,944$-$20.  Even for case {\sl (i)} when we
only consider atomic absorption lines in our computation of synthetic
spectra, we cannot see the extended but weak wings of the lines due to
the blending effects by other features.

From the observations of LP\,944$-$20, the pEW of the Li
resonance doublet in the UVES spectrum shown in Fig. \ref{__Liba} is
0.65 $\pm$ 0.05 \AA. Our pEW measurement is consistent with that (pEW =
0.53 $\pm$ 0.05 \AA) of Tinney (1998), suggesting little variability
in the overall strength of the line. We note that these pEWs are very
similar to the Li I measurements obtained for $\sigma$~Orionis members
(3 Myr) of related spectral types by Zapatero Osorio et al. (2002).
Using curves of growth measured by these authors for the young
M-dwarfs in $\sigma$ Orionis (see Fig. 16 in Zapatero Osorio et
al. 2002), we obtain a lithium abundance of log N(Li) = 3.2 $\pm$ 0.3
for LP\,944$-$20. Interestingly, the pEW of the Li I line computed for
log N(Li) = 3.25 and derived from the synthetic spectrum shown in
Fig. \ref{__Liba} (which accounts for TiO absorption) is 0.65~\AA,
 in agreement with the measurement of LP\,944$-$20.

Nevertheless, synthetic spectra are the most appropriate tool for 
a reliable lithium line analysis. Fits of our theoretical
spectra computed for the DUSTY model atmosphere 2000/4.5/0.0 are shown
in Fig.  \ref{__Liba};  the bottom panel of the Figure shows the
lithium line in larger scale. Theoretical spectra were computed for
several lithium abundances, from depletion by a factor of 10 up
to a rich content of log N(Li) = 3.5. The best fit is provided by
log N(Li)=3.25 $\pm$ 0.25 (see details in the bottom panel of Fig.
\ref{__Liba}). 
 LP\,944$-$20 seems to have preserved its original
lithium content.

It is worth noting that {\sl (i)} after appropriate broadening (see
Section \ref{__liobs}) the intensity and profiles of the Li I
resonance doublet can be well fit, and {\sl (ii)} the shape of TiO
bands around 670.8 nm cannot be reproduced with standard models. 
Theoretical spectra predict stronger TiO bands than those observed in
LP\,944$-$20 (we note that a change of 0.5 dex in gravity does not
have a major impact on the TiO bands in this wavelength range).
In the next section we attempt to improve the the poor fit to the TiO 
bands by adjusting our model atmosphere in a manner which seems physically 
justified.

Lithium subordinate lines at 812.6 and 610.3 nm can also be used for
Li abundance determination in atmospheres of some lithium-rich
late-type stars.  Moreover, any abundance derivation becomes more solid if
measurements are obtained from as many lithium lines and
lithium-bearing species as possible. 
We find the Li line at 610.3 nm ``sinks'' under the TiO bands 
in the LP\,944$-$20
spectrum. Only the 812.6 nm line can be used for the analysis if
reasonably good quality high-resolution spectra were available at
these wavelengths. The  CASPEC spectrum of LP\,944$-$20 does 
not comply
this requirement.  Therefore, in this paper our  abundance
analysis will be based on the fits to the observed Li I 670.8 nm line
profile.

\section{\label{semodel} Analysis with a semi-empirical model atmosphere}

In the framework of the standard model we have determined the
``cosmic'' lithium abundance [log N(Li)=3.25 $\pm$ 0.25] in the
atmosphere of LP\,944$-$20.  The same result was obtained from the
spectral analysis of the lithium resonance doublet profile and
the study of the line pEW. However, there remain a number of
problems. Our fit to the observed spectral energy distribution of
LP\,944$-$20 is not adequate in both the optical and infrared
regimes. A comparison with high resolution UVES spectra shows
significant differences in the shapes of computed and observed
spectra. Namely, the heads of TiO bands look ``smoothed'' in the
observed spectra. We find our fits obtained in the framework of
the standard model cannot be improved by changes of \Tef, log $g$~,
 metallicity or microturbulent velocity.

Generally speaking, these differences reduce the reliability of our
analysis, but on the other hand we know dusty effects can affect the
temperature structure and spectra of ultracool dwarfs (Tsuji et
al. 1996; Jones and Tsuji (1997); Allard et al. 2000; Burrows et
al. 2002b, and references therein). Therefore we construct the new
semi-empirical (SE) model atmosphere. With the new SE model atmosphere
we obtain better fits to the overall spectral energy distribution and
molecular features around lines of K, Rb and Li resonance
doublets. Finally, we  re-determine the lithium abundance with
the new model atmosphere.

\subsection{Semi-empirical model atmosphere}

We modified the DUSTY models as follows:
\begin{itemize}
\item To decrease the intensity of TiO bands in the spectral range
  600--800 nm, we suggest the complete absence of TiO above a certain
  height in the atmosphere of LP\,944$-$20. This can have at least
   two different physical explanations: {\sl (i)} the presence of
  a hot chromospheric-like region in the outermost layers of the
  dynamical atmosphere, or {\sl (ii)} the removal of Ti atoms 
  from gaseous species by condensation into dust particles in the
  upper atmosphere (see Pavlenko 1998a; Ferguson et al. 2005). In our
  SE model  we will explore the latter case, i.e., a depletion of
  TiO above a particular level in the atmosphere of LP\,944$-$20. Our
  lower boundary of the depleted TiO is located at the level
  $\tau_{\rm ross}$ = 10$^{-4}$. The choice of this point has some
  influence on  the absolute values of min $S$, however, it is
  not crucial for our results of lithium abundance determination.
\item To improve the fits over the wavelength range 0.8--0.9 and
  1.6--2.5 \mum, we implement additional quasicontinuum opacity
  (hereafter AqO) provided by dust particles. It is worth noting that
  dust particles must be present in atmosphere of LP\,944$-$20 due to
  the low effective temperature of the dwarf. Our knowledge about dust
  opacity in atmospheres of ultracool dwarfs is rather poor (see,
  however, review by Ferguson et al. 2005). Here we treat AqO as a
  scattering with power-law dependence on wavelength in the form
  $a_{\lambda} = a_0 \times ({\lambda_0/ \lambda)^{N}}$. We used
  $\lambda_0$ = 0.77 \mum as our reference wavelength. Parameter $N$
  varies in the range 0--4. A value of $N$ = 0 corresponds to the
  light scattering by submicron particles (``white scattering''), and
  $N$ = 4 describes Rayleigh scattering.
\item In our case, the ``dust opacity'' originates in the shell-like
  structures laying above the photosphere (clouds). It is worth noting
  that we tried different heights for the location of the ``dusty
  shell''. The best results are obtained for levels corresponding to
  $T_{\rm cr} \sim$ 2000 K,  coinciding with the models of Tsuji
  (2002) and Allard et al.  (2001).  These authors refer to $T_{\rm
  cr}$ as a temperature related to the ``gas-dust'' phase transition.
\end{itemize}

\begin{figure}   
\includegraphics [width=88mm]{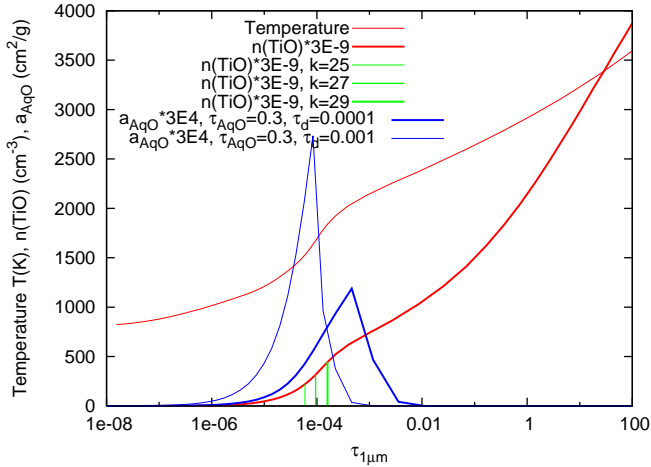}
\caption[]{\label{__mtio} Structures of two semi-empirical model
atmospheres from our grid.}
\end{figure}

\begin{figure}   
\includegraphics [width=88mm]{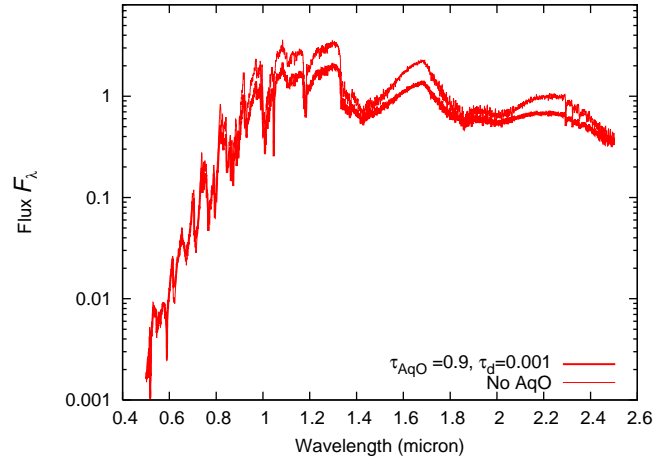}
\caption[]{\label{__sedi} Comparison of fluxes computed with the DUSTY
2000/4.5/0.0 model atmosphere for two different cases: the dust-free
atmosphere (no AqO) and the SE model (\TAO = 0.9, \TSL = 0.001, $k$
= 25). See also Fig. 2 of Burrows et al. (2006).}
\end{figure}

Our scattering dusty cloud structure was modelled by two parameters:
thickness of cloud, \TAO, and the location of the maximum of opacity,
\TSL. The dusty opacity decreases inward and outwards in the
atmosphere with respect to the \TSL point (see Fig. \ref{__mtio}) for
 physical reasons: pressure and density are reduced outwards 
while temperature increases inwards. These factors naturally
decrease the density of the dusty particles at lower and larger
optical depths. We vary \TAO and \TSL across the ranges 0--1 and
0.001--0.0001, respectively. Note  that different locations of the
scattering layer correspond in our case to different depths in the
atmosphere in which $T$ = $T_{cr}$.   Burrows et al. (2006) have
recently proposed a similar model of the dusty cloud with a flat part
of the dust distribution located at T $\sim$ 2300 K.

\subsection{Fits to the observed spectral energy distribution}

We have computed a grid of theoretical spectra  adopting our SE
model atmospheres built with different parameters of \TAO, \TSL, and
$k$  ($k$ stands for different layers within the atmosphere). A
comparison of computed optical and near-infrared SEDs for the SE
model atmospheres with AqO (\TAO=0.9, \TSL=0.001) and without AqO is
shown in Fig. \ref{__sedi}. As we see from the comparison, our AqO
affects the spectral regions where fluxes are formed in the layers
below the level $\tau <$ \TAO. Strong absorption features formed in
the optical spectrum are affected rather marginally.  In general,
the AqO makes the computed spectra appear ``shallower'' than models
without AqO, i.e., in agreement with the observations of
LP\,944$-$20.

\begin{figure}   
\includegraphics [width=88mm]{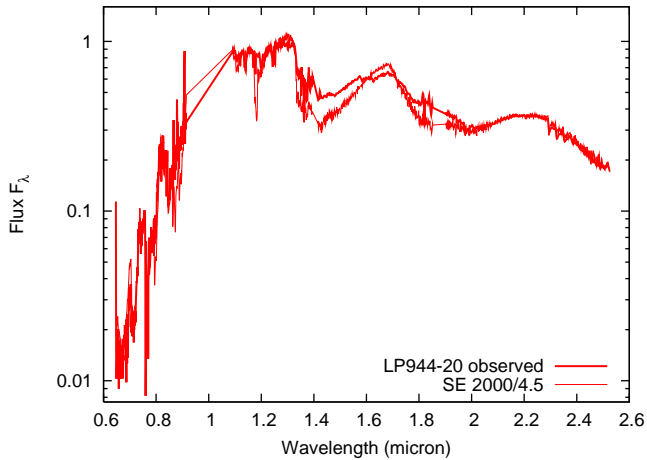}
\caption[]{\label{__SEDD} Fit of the theoretical fluxes computed with
a SE model atmosphere (2000/4.0/0.0, \TAO = 0.9, \TSL = 0.001, $k$ =
25) to the observed spectral energy distribution of LP\,944$-$20. Note
the better fit as compared to Fig. \ref{__fitA0}.}
\end{figure}

Theoretical SEDs computed for SE model atmospheres were fit to the
observed fluxes of LP\,944$-$20 following the procedure described in
section \ref{eq}. It is worth noting  the following: {\sl (i)}
the minimisation procedure of fits to observed spectra provides lower
$S$ for theoretical spectra computed with $N$ = 0  (this suggests
that the dusty particles present in the modelled dusty cloud structure
of LP\,944$-$20 are of submicron size), and {\sl (ii)} visual
comparison of Figs. \ref{__fitA0} and \ref{__SEDD} shows that we get a
better fit to the observed SED with theoretical spectra computed with
AqO. Analysis of the results of the numerical procedure of
minimisation yields $S$ = 1.4$\times$10$^{-6}$ and
3.9$\times$10$^{-7}$ for the fits shown in Figs. \ref{__fitA0} 
(standard approach) and \ref{__SEDD} (SE model) respectively. We note
that while our SE computations decrease the absolute values of min
$S$, the general shape of the min $S$ dependence on \Tef (like the one
shown in Fig. \ref{__As}) is not significantly changed. SE models of
\Tef = 2000 K provide reasonable fits to the observed data of
LP\,944$-$20, indicating that our results are stable.

\subsection{K I and Rb I resonance lines}

\begin{figure}    
\includegraphics [width=88mm]{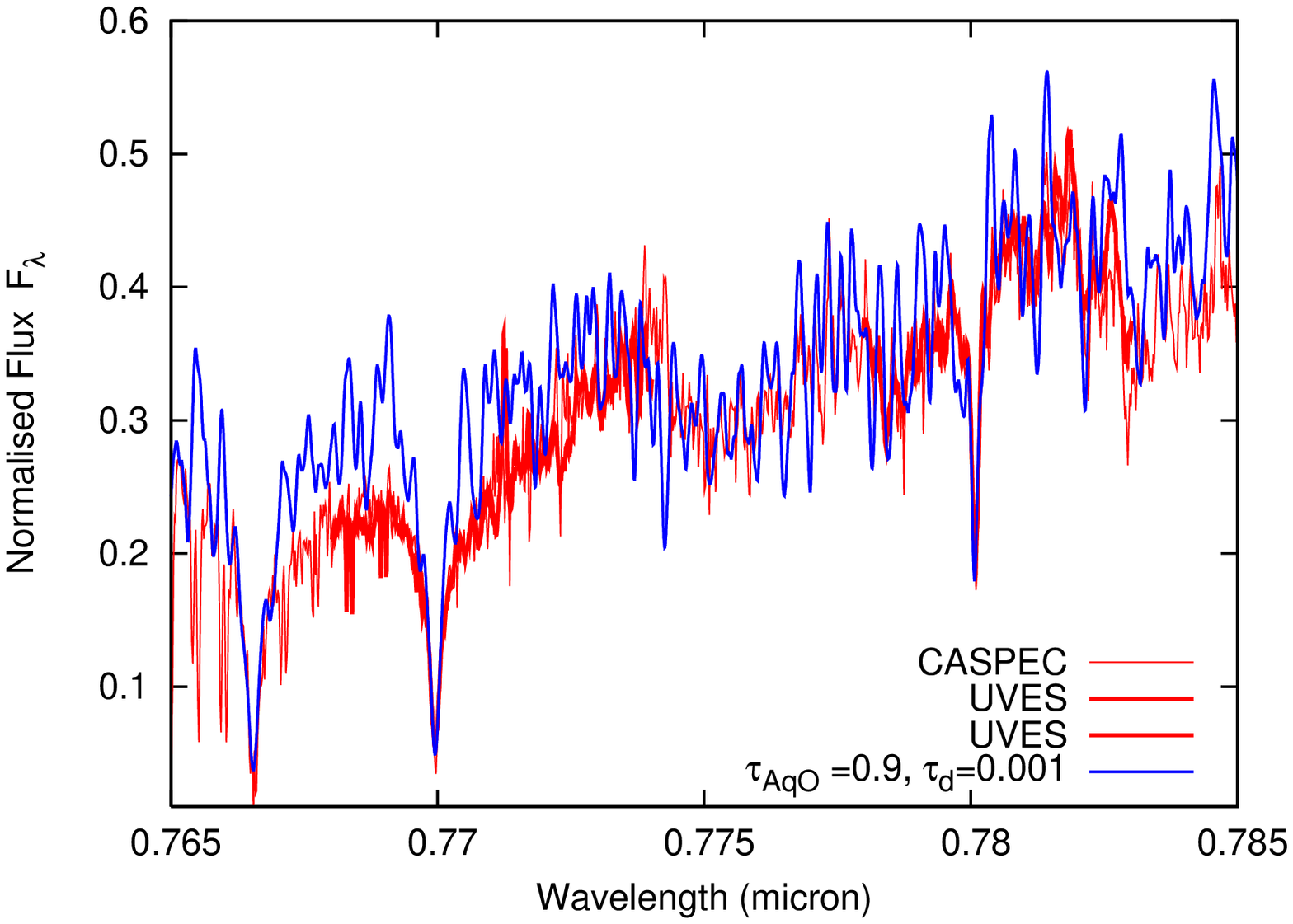}
\includegraphics [width=88mm]{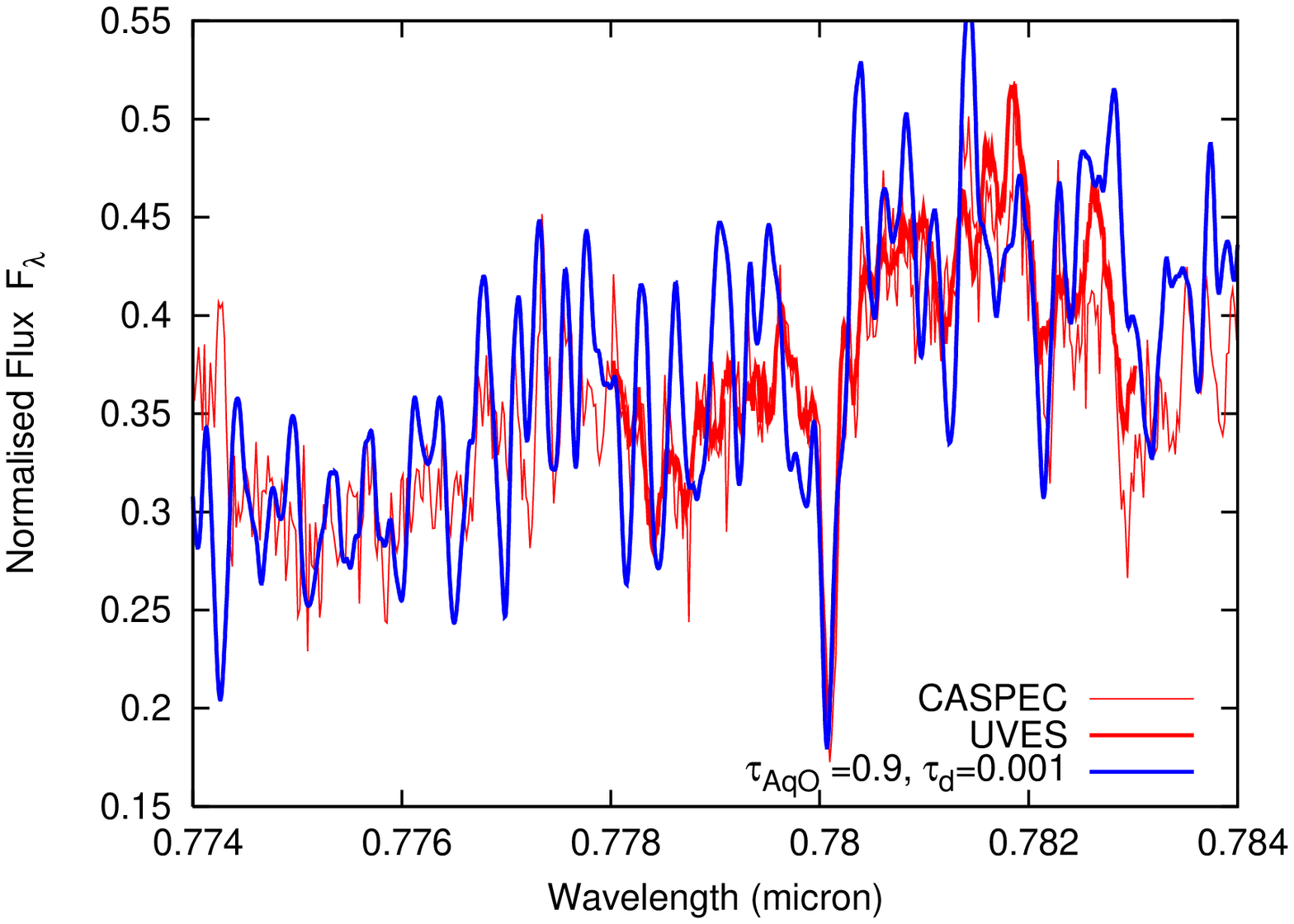}
\caption[]{\label{__KRbba} Fits of the synthetic fluxes computed with a
SE model atmosphere (2000/4.5/0.0, \TAO = 0.9, \TSL = 0.001, $k$ =25)
to the LP\,944$-$20 observed spectrum around K I and Rb I lines. The
bottom panel shown an enlargement around the Rb I line.}
\end{figure}

Fits of theoretical spectra computed for SE model atmosphere to the
spectral region of K I and Rb I line are shown in Fig.
\ref{__KRbba}. We get even better fits to K I and Rb I resonance lines
 and to TiO molecular features in terms of intensity in
comparison to the dusty free case (see Fig. \ref{__KRbaa}).

\subsection{\label{__IRK} Near-infrared doublet of K I}

\begin{figure}    
\includegraphics [width=88mm]{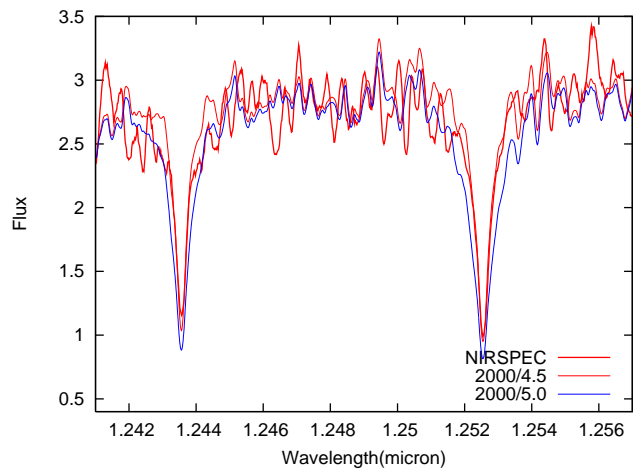}
\caption[]{\label{__IRb} Fit of computed SE models (2000/4.5/0.0 and
2000/5.0/0.0, \TAO = 0.9, \TSL = 0.001, $k$ =25) to the LP\,944$-$20
observed infrared K I subordinate doublet region. The core of the K I
lines are better reproduced than in Fig. \ref{__IRa}. Note that log
$g$ = 4.5 provides a better match to the profiles of these
``gravity-sensitive'' atomic lines.}
\end{figure}

 The NIRSPEC data of LP\,944$-$20 is compared to our synthetic
spectrum computed with the SE model atmosphere in Fig.  \ref{__IRb}.
In general, our SE model atmosphere does not have a major impact on
the atomic and molecular details of these wavelengths: the fit of the
molecular features around K I lines are of the same quality and the
profiles of the K I lines are reproduced as in the AqO-free case (see
Fig. \ref{__IRa}).  However, we note that the fits to the
observed cores of K I are  improved in the AqO model. And again,
we see that models of log $g$~ = 4.5 provide a much better fit  to
the atomic lines than higher gravity atmospheres.

\subsection{Li I resonance lines}

\begin{figure}    
\includegraphics [width=88mm]{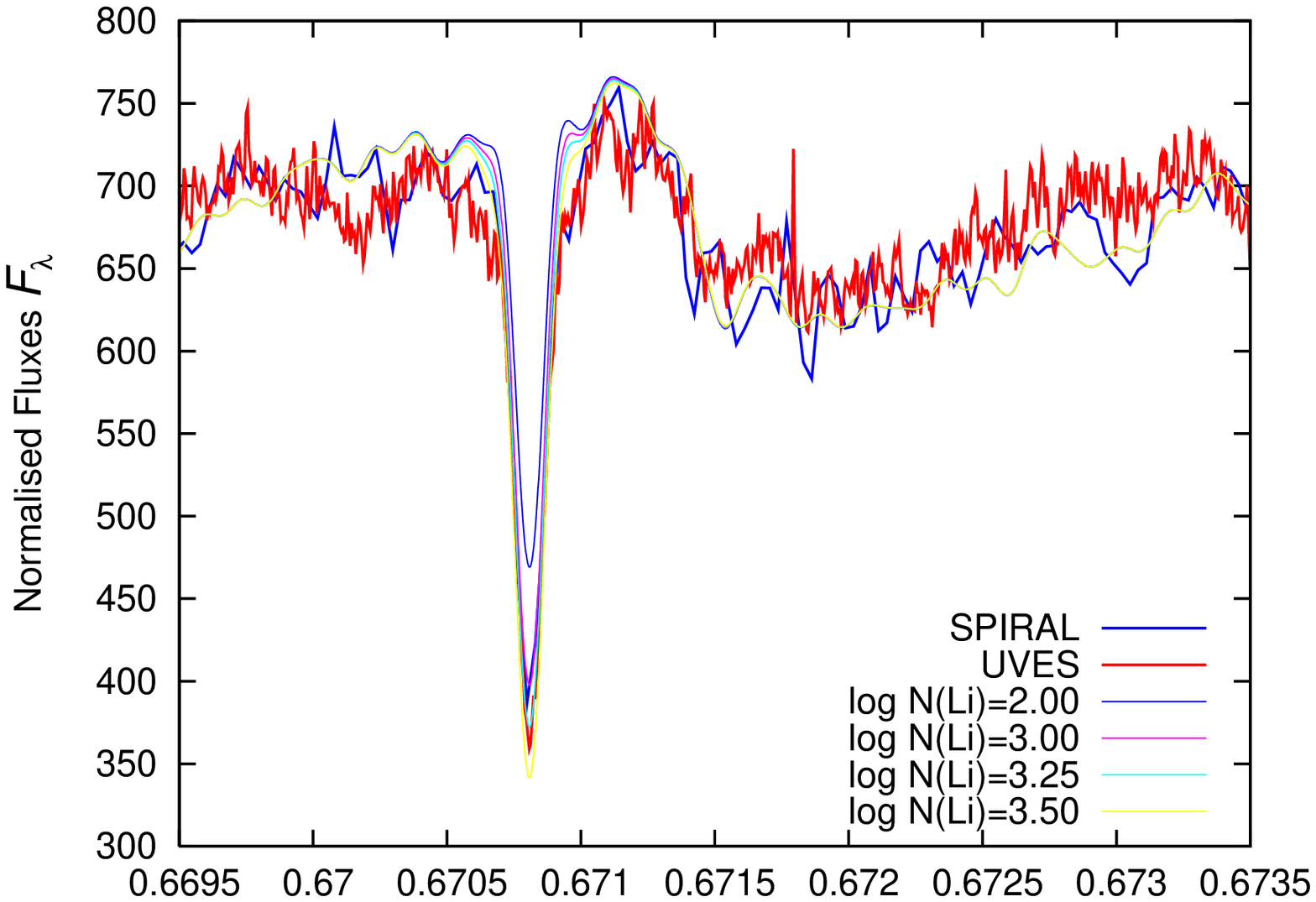}
\includegraphics [width=88mm]{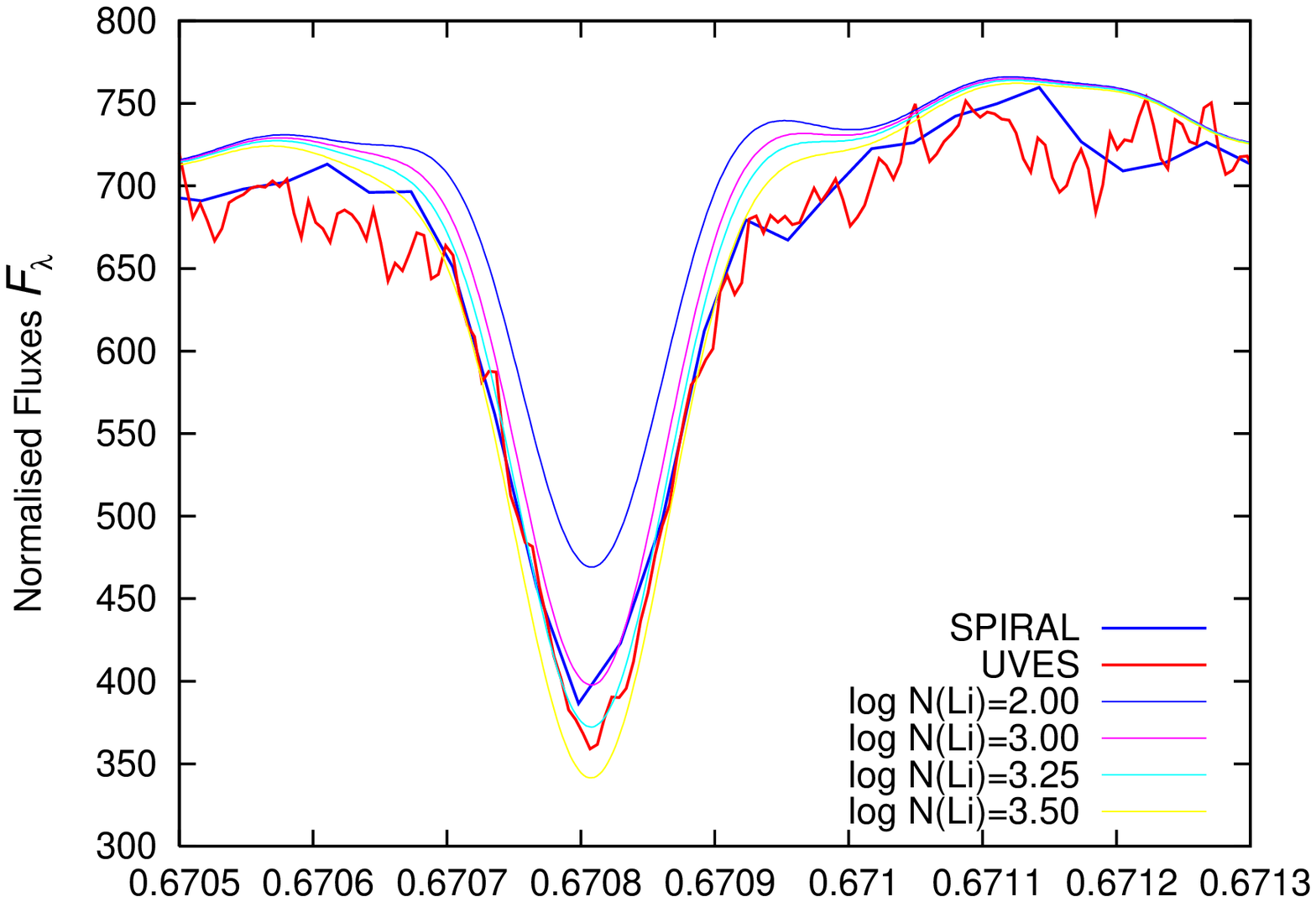}
\caption[]{\label{__Lica} Fits of the theoretical fluxes around Li
  670.8 nm resonance doublet computed with SE model atmosphere
  (2000/4.5/0.0, \TAO = 0.9, \TSL = 0.001, $k$ = 25) and different
  lithium abundances to the observed spectrum of LP\,944$-$20. The
  bottom panel shows an enlargement around the resonance line. Note
  the better reproduction of the shape and intensity of the TiO
  absorption as compared to Fig. \ref{__Liba}.}
\end{figure}

 The SPIRAL and UVES spectra of LP\,944$-$20 around Li I 670.8 nm
are compared to SE models in Fig. \ref{__Lica}. These models have been
computed for different lithium abundances, from log N(Li) = 2.0 up to
3.5. The fits to the resonance line profile of Li I are of similar
quality to those depicted in Fig. \ref{__Liba} (``standard
models''). However, the TiO absorption fit and the overall fit of the
spectral region across the Li resonance doublet is significantly
improved in the SE models. 

We carry out our lithium abundance determination for the SE model
atmosphere.  As seen from the bottom panel of Fig. \ref{__Lica},
the model computed for log N(Li) = 3.25 provides the best match to the
observations. Interestingly, our SE model does not significantly affect
the lithium abundance determination for
LP\,944$-$20. Observed and theoretical spectra have been normalised
over a broad range of wavelengths across the lithium region because
the synthetic data provide a reasonable reproduction of the TiO
absorption. We have adopted an error bar of $\pm$0.25 dex in our
lithium abundance determination; this takes into account the quality
of the observed spectra, \Tef changes of $\pm$200 K, and slight
modifications in the normalisation of the spectra. From our
computations and at the 4 $\sigma$ confidence level we can discard
lithium depletion factors larger than 10 in the atmosphere of
LP\,944$-$20.

\section{Discussion}

Our main result has been obtained from the  spectral analysis of
the high-resolution profile of the Li I doublet resonance line  at
670.8 nm using \Tef = 2000 K (\Tef = 2040 K is derived for
LP\,944$-$20 from photometric and astrometric considerations), log $g$
= 4.5, solar metallicity, and different atmospheric lithium contents
from depletion of about a factor of 10 up to super-cosmic abundance.
We also model other alkali lines of K I and Rb I located in other 
optical and near-infrared spectral regions  as a test of
consistency of our results.  We note that profiles of the alkali
lines obtained  on different occasions agree quite well in
intensities and widths. Some discrepancies may be explained by the
signal-to-noise differences or variability of dust shell within the
atmosphere of LP\,944$-$20.

Firstly, we reproduce the overall spectral distribution and profiles of
spectral lines of strong absorption lines of rubidium and
potassium. Second, we find the lithium abundance in the atmosphere of
LP\,944$-$20 by fitting a DUSTY model atmosphere to the Li 670.8 nm
line profile in the observed spectrum. We find that the fit to the
observed fluxes provides a cosmic value of lithium abundance and that
analysis of the the observed pEW also yields the same log N(Li)
= 3.25 $\pm$ 0.25 dex.

Tinney (1998) determined a lithium abundance of log
N(Li)=0.0 $\pm$ 0.5 from analysis of the EW of the lithium
resonance doublet. We can reproduce Tinney's (1998) 
result if we use pEW as EW (see 
Section \ref{__ll}), as Tinney was forced to do
by the then available atmospheric models.
Indeed, only pEW of atomic lines can be
measured in respect to the background formed by the haze of TiO lines
(Pavlenko 1997, 1997a; Zapatero Osorio et al. 2002). The pEW values
depend on the strength of the background molecular bands of TiO,
on the spectral resolution of the data, and on other broadening
parameters (see Mart\'\i n et al. (2005) for an example of how
rotation affects the equivalent widths of K I resonance lines).
Reducing the TiO absorption in the region across the 670.8 nm line 
increases the relative strength of the computed Li lines
and the formally determined lithium abundance decreases.

\begin{figure}   
\begin{center}
\includegraphics [width=88mm]{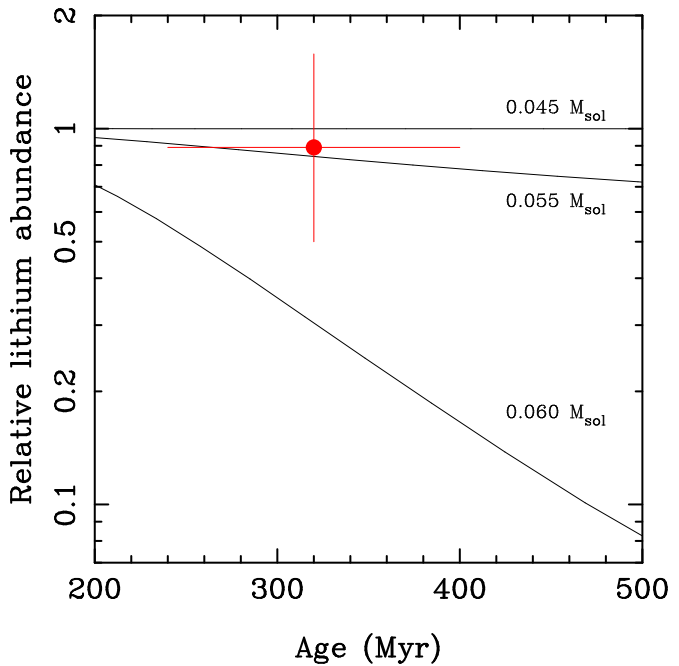}
\end{center}
\caption[]{\label{__evol} Location of LP\,944$-$20 on the lithium
depletion curves by Baraffe et al. (1998). We have adopted the age
determination of Ribas (2003).}
\end{figure}

\begin{figure}   
\begin{center}
\includegraphics [width=88mm]{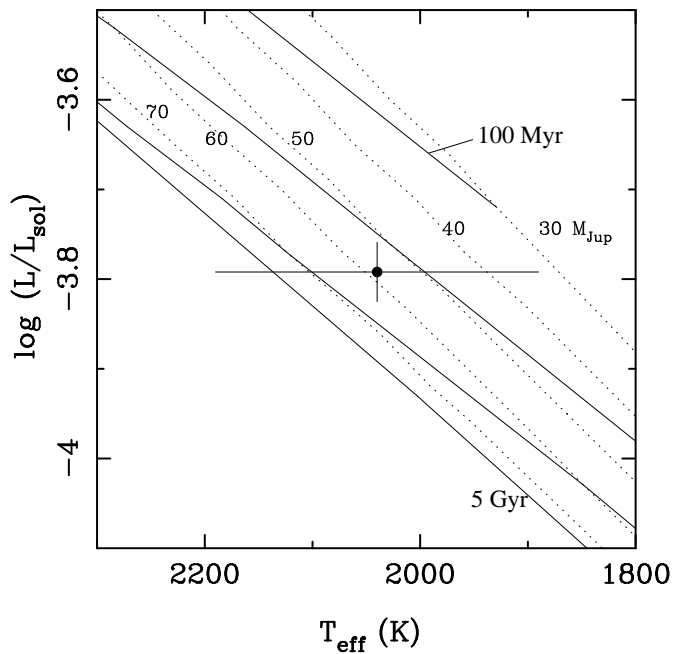}
\end{center}
\caption[]{\label{__evol_hrd} Location of LP\,944$-$20 in the H-R
diagram and comparison with the evolutionary models of Baraffe et
al. (1998). Isochrones (solid lines) correspond to the following ages:
from top to bottom, 100 Myr, 300 Myr, 800 Myr, and 5 Gyr. Evolutionary
tracks (dotted lines) have masses labelled in Jovian units (1
$M_\odot$ $\sim$ 1000 $M_{\rm Jup}$).}
\end{figure}

It is worth noting that we get some problems with fits of theoretical
spectra computed for the conventional, i.e., self-consistent, model
atmosphere to the observed spectral energy distribution  of
LP\,944$-$20. The shape and intensity of the TiO bands around the
resonance line of lithium are not well matched by ``standard''
models.  Generally speaking,  because of the low surface
temperature and activity properties of LP\,944$-$20, it might be
surprising to obtain good fits for conventional models, given that
LP\,944$-$20 provides substantial evidence of stellar
activity phenomena in its atmosphere (H$\alpha$ and radio emission,
strong X-ray flares).  Our knowledge of these processes is far from
complete even in the case of the Sun.  Indeed, the presence of
these phenomena is clear evidence of the vertical stratification of
the LP\,944$-$20 atmosphere.  On the other hand, it is plausible
that TiO is absent in the upper atmosphere of LP\,944$-$20 due to
depletion of Ti atoms into dust particles and/or complete dissociation
of TiO due to higher temperature there.

To investigate the impact of different factors on our main result 
and in order to provide a better fit to the observed spectra of
LP\,944$-$20, we adopt a more sophisticated model atmosphere. Namely,
we suggest the presence of clouds at the level above the photosphere
and complete depletion of TiO in the outermost layers of the
atmosphere of LP\,944$-$20. We treat the opacity provided by the dusty
clouds as scattering with power dependence on wavelength. From the
comparison of observed and computed spectra we determine parameters of
 scattering clouds. The temperature-pressure dependence in our
model atmosphere was taken from the DUSTY model atmosphere  of the
Lyon group, i.e., our SE model is not self-consistent.  Even in the
Sun (e.g., HOLMU, VLC, MAKKLE), the use of SE model atmospheres is
well acknowledged (see references in Gibson
1970). Semi-empirical model atmospheres are used to fit some observed
features using theoretical spectra computed with modified model
atmospheres.

In our case the appearance of a dusty shell in the SE model atmosphere
and/or depletion of TiO should change the temperature structure of the
outermost atmospheric layers.  With our SE models we obtain
considerably better fits between observational and synthetic spectra
in terms of the TiO bands, alkali lines and SED.  From fits to
the Li resonance doublet we  conclude that LP\,944$-$20 has likely
preserved all its originial lithium content.

We also conclude that our lithium abundance shows a rather weak
dependence on the input parameters of the models, despite the fact
that our log N(Li) determination is carried out in the frame of local
thermodynamical equilibrium (LTE). Moreover, all the alkali lines
fitted in our paper were computed in the frame of LTE.  Fortunately,
the alkali lines considered in this work are rather insensitive to
chromospheric-like features (CLF) because processes of their formation
are mainly controlled by photoionisation (Thomas 1959). By definition,
these lines should show a rather weak dependence on temperature
structure of the outermost layers of atmosphere. Indeed, direct NLTE
modelling of formation of lithium lines in the atmospheres of
ultracool dwarfs with CLF shows their rather weak response to
temperature inversions, in contrast to TiO lines (Pavlenko
1998).  Lithium lines can be affected only in the case of the
strongest CLF producing additional flux in the blue part of the
spectra, but that is unlikely in LP\,944$-$20 given its age and level
of H$\alpha$ and X-ray activity.

Finally, comparison of our revised lithium abundance for
LP\,944$-$20 with the calculations of lithium depletion by Baraffe et
al. (1998) shows a good agreement for a mass below 0.057 $M_\odot$ and
an age of 320$\pm$80 Myr (see Fig. \ref{__evol}). 
This  result is marginally consistent with the earlier estimate of
Tinney (1998), who derived a mass in the range 0.056 -- 0.064 $M_\odot$.
Using the same  evolutionary models, we show that
this age and  a mass $\le$0.057 $M_\odot$ are also consistent
with the location of this brown dwarf in the H-R diagram
(Fig. \ref{__evol_hrd}).  Thus, our analysis supports the  young
age of a few hundred million years of LP\,944$-$20 as suggested by
Ribas (2003), and indicates a lower mass  than previously reported
in the literature.

\section{Acknowledgments}

YP's studies are
supported the Royal
Society and  Leverhulme Trust.
EM acknowledges support from
NSF research grant AST 02-05862, the Michelson Science Center,  
and Spanish MEC grant AYA2005-06453.  
We thank Isabelle Baraffe for providing evolutionary models 
upon our request. We thank Chris Tinney for the helpful comments. 
We are grateful to the user support group of the VLT. 
The W.M. Keck Observatory is operated as a scientific 
partnership between the California Institute of Technology, 
the University of California, and NASA. The Observatory was made 
possible by the generous financial support of the W.M. Keck Foundation. 
The authors extend special thanks to those of Hawaiian ancestry on 
whose sacred mountain we are privileged to be guests.   
This research has made
use of the SIMBAD database, operated at CDS, Strasbourg, France.


\end{document}